\documentclass{emulateapj}

\usepackage{epsf}
\usepackage{relsize}
\usepackage{verbatim}

\newcommand{\ie}{\emph{i.e.,} }
\newcommand{\eg}{\emph{e.g.,} }
\newcommand{\pr}{\partial}
\newcommand{\km}{\mbox{ km}}
\newcommand{\se}{\mbox{ s}}
\newcommand{\pc}{\mbox{ pc}}
\newcommand{\fin}{\mbox{ .}}
\newcommand{\coma}{\mbox{ ,}}
\newcommand{\const}{\mbox{const}}
\newcommand{\lrgspc}{\,\,\,\,\,\,\,\,\,}

\newcommand{\rmvsmlspc}{\!\!\!\!}

\newcommand{\myw}{w}
\newcommand{\myp}{p}
\newcommand{\myq}{q}
\newcommand{\myk}{k}
\newcommand{\myh}{h}
\newcommand{\myT}{\mu}
\newcommand{\mytau}{\mu}
\newcommand{\myr}{\zeta}
\newcommand{\myg}{\xi}
\newcommand{\myrho}{\rho}
\newcommand{\myx}{x}

\newcommand{\mydelta}{\delta}

\begin{document}

\title{Analytic study of mass segregation around a massive black hole}

\author{Uri Keshet\altaffilmark{1}\altaffilmark{2}\altaffilmark{5}, Clovis Hopman\altaffilmark{3}, and Tal Alexander\altaffilmark{4}}

\altaffiltext{1}{Center for Astrophysics, 60 Garden Street, Cambridge, MA 02138, USA}

\altaffiltext{2}{Institute for Advanced Study, Einstein Drive, Princeton,
NJ, 08540, USA}

\altaffiltext{3}{Leiden University, Leiden Observatory, P.O. Box 9513, NL-2300 RA Leiden, The Netherlands }

\altaffiltext{4}{Faculty of Physics, Weizmann Institute of Science, P.O. Box 26, Rehovot 76100, Israel}

\altaffiltext{5}{Einstein fellow}

\date{\today}

\begin{abstract}
We analyze the distribution of stars of arbitrary mass function $\myg (m)$ around a massive black hole (MBH). Unless $\myg $ is strongly dominated by light stars, the steady-state distribution function approaches a power-law in  specific energy $\myx\equiv -E/m\sigma^2<\myx_{max}$ with index $p=m/4M_0$,
where $E$ is the energy, $\sigma$ is the typical velocity dispersion of unbound
stars, and $M_0$ is the mass averaged over $m \myg  \myx_{max}^p$. For light-dominated $\myg $, $p$ can grow as large as $3/2$ -- much steeper than previously thought.
A simple prescription for the stellar density profile around MBHs is provided. We illustrate our results by applying them to stars around the MBH in the Milky Way.
\end{abstract}


\maketitle

\section{Introduction}

A massive black hole (MBH) of mass $M_\bullet$ dominates the dynamics
of stars within its radius of influence
$r_h=GM_\bullet/\sigma^2=2.3 (M_\bullet/3\times 10^6
M_\odot)(\sigma/75\km\se^{-1})^{-2}\pc$, where $\sigma$ is the typical
star velocity dispersion at $r\gtrsim r_h$ and values quoted
correspond to the Milky Way center \citep{Ale05}. The steady-state
distribution function (DF) of such stars, first derived (for simple, discrete
stellar mass functions, MFs) by Bahcall and Wolf (1976, 1977; henceforth
BW76,77) is useful in the study of galactic centers and
possibly globular clusters. In the simple case where
all stars have equal mass, BW76 showed that an $n\propto r^{-7/4}$
density cusp forms; a somewhat flatter cusp was subsequently identified around SgA*
\citep{Ale99a,Gen03a,Sch07}.

Significant mass segregation is expected for realistic MFs
\citep[][and references therein]{Mir00,Sch07,OlearyEtAl08}, as dynamical friction slows heavy stars that sink towards the MBH while light stars are pushed outwards.
This effect
is essential in modeling various physical processes, such as tidal
disruption \citep{Lig77, Mag99, Sye99} and gravitational wave emission
\citep{HopmanAlexander06, Fre06, Hop07a},
and may be used for example to test if intermediate-mass black holes exist in globular clusters \citep{Gill08}.
Recently, \citet[][henceforth AH09]{AlexanderHopman08_StrongSegregation} showed strong mass segregation in  the limit of a strongly bimodal MF when the mass ratio is large and massive stars are rare.

In spite of extensive numerical studies of mass segregation around a MBH
(AH09 and references therein),
analytical modeling (BW77) is limited to simple,
discrete, and extreme MFs where the most massive species constitutes more
than half of the star population. Most results are limited to few stellar
species and to a narrow mass range. In this \emph{letter} we analytically
study the steady-state single-star DF $f$ within $r<r_h$, for an arbitrary,
continuous MF. We confirm our results numerically
for a wide range of MFs.

Following BW76,77, we assume (i) spherical spatial symmetry; (ii) isotropic velocities; (iii) Keplerian orbits; (iv) binaries are negligible;
(v) small angle, uncorrelated, local scatterings; (vi) loss cone effects can be neglected; (vii) isothermal distribution of unbound stars with temperature $\myT \sigma^2$; (viii) stars are destroyed when their specific energy drops below a threshold $x\equiv -E/m\sigma^2 = x_{max}$,
with $E$ the energy.

\section{Equal mass stars}
\label{sec:EqualMassStars}
Consider the  case where all stars have mass $m$. The dimensionless diffusion rate of stars through $x$ is (BW76)
\begin{equation}
\label{eq:BW_Integral} Q(x) = \int_{-\infty}^{x_{max}}
\frac{f(x)f'(x')-f'(x)f(x')}{\left[\mbox{Max}(x,x')\right]^{3/2}}\,dx' = \const \fin
\end{equation}
The boundary conditions are $f(x<0)=e^{x m/\myT}$ and $f(x>x_{max})=0$.
Eq.~(\ref{eq:BW_Integral}) with
these boundary conditions uniquely determines $f$ and $Q$ as shown below.
The spatial number density of stars is given by (BW76)
\begin{equation} \label{eq:spatial_density}
n(r) \propto \int_{-\infty}^{r_h/r} \,dx
f(x)\sqrt{\frac{r_h}{r}-x} \fin
\end{equation}

Eq.~(\ref{eq:BW_Integral}) describes the balance between diffusion towards
the MBH and replenishment by new stars.
BW76 found that $Q$ is essentially determined by the
diffusion rate at a bottleneck near $x_{max}$, where the replenishment
rate diminishes. As $Q\simeq 8/x_{max}$ is typically small, one can
approximately solve Eq.~(\ref{eq:BW_Integral}) by setting $Q=0$.

We convert Eq.~(\ref{eq:BW_Integral}) to an ordinary differential equation (ODE) by
repeated operations of differentiation with respect to $x$ and isolation
of integral terms. This yields a nonlinear, fourth-order ODE for $f$,
\begin{equation}
\label{eq:DEsingle} \myq f^2\left(B+4\myp A\right) = x^{3/2} Q C \fin
\end{equation}
Here we defined the local power-law index of $f$, $\myp \equiv d\ln f/d\ln x=x\myw$, the local power-law index of $\myw\equiv (\ln f)'$, $\myq
\equiv d\ln \myw/d\ln x$, and the operators
\begin{eqnarray}
\label{eq:ABC_defs}
A(f,x) & \equiv & 4\myq^2-5\myq-2 x^2\myw''/\myw \coma \\
B(f,x) & \equiv &
5\myq-12x^3\myw''^2/(\myw'\myw)+8x^3\myw^{(3)}/\myw
\nonumber
\coma \\
C(f,x) & \equiv & \frac{\myp A^2 D}{3}\left[ \frac{B}{A^2}
-\frac{4}{D} + \frac{3}{\myq}\right. \nonumber \\
& &  \lrgspc \lrgspc \left. + \frac{2}{A}\left(
10+2\myq-2\myp+\frac{5}{D} \right) \right] \nonumber \coma
\end{eqnarray}
where $D(f,x)\equiv \myq+\myp-5/2$. The boundary conditions at $x=0$ fix
$f$ and $f'$ there, so the non-exponential solution $f(x)$ is completely determined for given $Q$.
A unique value of $Q$ guarantees that $f$
vanishes (for the first time) at $x_{max}$, proving the uniqueness
of the steady-state solution (an ``exercise left for the reader'' by BW76).

For a power-law DF $f =f_0 x^{\myp_0}$, Eq.~(\ref{eq:DEsingle})
becomes
\begin{equation} \label{eq:fixed_p0}
3f_0^2 \left(\myp_0-1/4\right) = Q x^{3/2-2\myp_0} \myp_0 \left(
\myp_0-1 \right) \left( \myp_0-5/2 \right) \fin
\end{equation}
This compactly reproduces the BW76 results: although a finite,
energy-independent flow requires $\myp_0=3/4$
\citep[][implying a flow of stars away from the MBH, see BW76]{Pee72}, a
steady-state with $\myp_0=1/4$ is closer to the actual DF
because $Q$ is negligibly small.
Note that the power-law assumption becomes inconsistent at high
energies, where $Q x^{3/2-2\myp_0}\propto x$ is large.

Henceforth we assume $Q=0$, justified for single mass stars if $x_{max}\gg 1$. An exponential DF with $\myq=0$ solves Eq.~(\ref{eq:DEsingle}), but does not satisfy the boundary conditions. For $q\neq0$, Eq.~(\ref{eq:DEsingle}) becomes a third order nonlinear ODE,
\begin{equation}
\label{eq:ODEF0} B+4\myp A=0 \coma
\end{equation}
with general solution of the form
\begin{equation}
\label{eq:solGeneral} \myw=\frac{c_1}{x^{-3/2}-c_2} +c_3 + c_1 c_3
W(x;c_1,c_2,c_3) \coma
\end{equation}
with $c_1,c_2,c_3$ constants.

Useful results can be deduced directly from Eqs.~(\ref{eq:ODEF0}) and
(\ref{eq:solGeneral}), without determining the analytic properties of $W$.
When $p$ is much smaller (larger) than $1/4$, Eq.~(\ref{eq:ODEF0}) becomes
approximately $B\simeq 0$ ($A\simeq 0$), which corresponds to taking $W$
(taking $c_3$) to zero in Eq.~(\ref{eq:solGeneral}). In this case $f(x\gg1) \propto \exp[\int^x w(x')dx']$ is either nearly constant, or (super-) exponential in $x$.
In order to
maintain reasonable stellar densities, there must be an approximate
balance between the two terms, implying that $p\simeq 1/4$ far from the
boundaries. Note that for $p=1/4$, the six terms of $A$ and $B$ in
Eq.~(\ref{eq:ABC_defs}) precisely cancel in pairs.

\begin{figure}[h]
\centerline{\epsfxsize=7cm \epsfbox{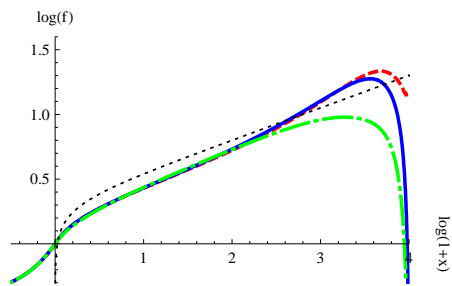}} \caption{
DF of equal mass stars around a MBH with $x_{max}=10^4$ and
$\myT/m=1$, according to (i) exact ($Q\simeq 8\times 10^{-4}$, solid) and
(ii) approximate ($Q=0$, dashed) numerical solutions of
Eq.~(\ref{eq:BW_Integral}), (iii) Eq.~(\ref{eq:approx_single}) (dash-dotted),
and (iv) $f\sim 2x^{1/4}$ (dotted). Logarithms are base 10 and convergence is better than $1\%$, henceforth. }
\label{fig:approx_single}
\end{figure}

For small $x$, where $0\leq\myp=x \myw\ll1$, Eq.~(\ref{eq:ODEF0}) becomes
$B\simeq 0$; this corresponds to Eq.~(\ref{eq:solGeneral}) with $W\to0$,
and $c_3=m/\myT$ according to the boundary conditions at $x=0$. At
intermediate energies where approximately $f\propto x^{\myp_0}$,
Eq.~(\ref{eq:DEsingle}) becomes $-12w''^2+8w^{(3)}w'\simeq0$, with all
other terms smaller by factors $\gtrsim 5$ ($\gtrsim 10$) for
$|\myp_0|<1/2$ ($|\myp_0|<1/4$), so
$\myw\simeq\myp_0(x+x_1)^{-1}+x_2^{-1}$. At high energies near the
disruption energy, $\myp$ is large and negative, Eq.~(\ref{eq:ODEF0})
becomes $A\simeq0$ and so $c_3=0$; the boundary condition at $x_{max}$
then yields $c_2=c_1/(3n)=x_{max}^{-3/2}$ with $n\in\mathbb{N}$. Smoothly
combining these approximate solutions yields an approximation for the equal mass case,
\begin{equation} \label{eq:approx_single}
f_{eq} \simeq \left\{
\begin{array}{ll}
\left(1+\frac{x m}{\myp_0 \myT} \right)^{\myp_0} e^{x/x_2} & \mbox{if $x<x_{ta}$;}
\\
\begin{array}{ll}
& \rmvsmlspc \widetilde{f} \left[ \frac{(\sqrt{x_{max}}-\sqrt{x})^2} {x_{max}+\sqrt{x_{max}
x}
+ x} \right]^n \exp \left[ \frac{3nx}{x_{max}} \right. \\
& \rmvsmlspc \left. +2n\sqrt{3} \arctan\left( \frac{1}{\sqrt{3}}
+\sqrt{\frac{4x}{3x_{max}}} \right) \right]
\end{array}
& \mbox{if $x>x_{ta}$,}
\end{array}
\right.
\end{equation}
with constants $x_2,x_{ta}$ and $\widetilde{f}$ determined for example by continuity of $f$,
$3nx_{max}^{-5/2}/(x^{-3/2}-x_{max}^{-3/2})=\myp_0x^{-1}+x_2^{-1}$ and its
first derivative evaluated at the turn-around energy $x_{ta}$. The results
depend weakly on $n$; it is natural to fix $n=1$. The previous paragraph
suggests that $\myp_0\to1/4$ as $x_{max}$ increases.
Numerically solving Eq.~(\ref{eq:BW_Integral}) yields $\myp_0=(1/4)\myrho(\myx_{max})$, where the correction factor $\myrho\sim 1$ approaches unity as $x_{max}$ becomes large, \eg $\myrho(10^4)=1.18$ and $\myrho(10^8)=1.08$. It also depends weakly on the
energy range used to measure $p$ (here $x_{max}^{1/3}<x<x_{max}^{1/2}$, henceforth).
Figure \ref{fig:approx_single} illustrates the approximation.

\section{Continuous Mass Function}
Consider stars with variable mass $m$ in some range $M_L<m<M_H$.
We generalize the discrete, multiple-mass version of Eq.~(\ref{eq:BW_Integral}) (BW77) to the continuum limit,
\begin{eqnarray}
\label{eq:BW_masses} Q(m) & = & \int_{M_L}^{M_H} m'\,dm' \int_{-\infty}^{\myx_{max}}
d\myx' \, \left[\mbox{Max}(\myx,\myx')\right]^{-3/2}
\\
& & \times \left[m f(\myx,m)\pr_{\myx'} f(\myx',m') - m'
\pr_{\myx}f(\myx,m)f(\myx',m') \right] \coma \nonumber
\end{eqnarray}
where $f$ is now the DF in $\myx-m$ phase space.
Assuming unbound stars with MF $\myg (m)$, $f(\myx<0,m)=\myg (m)e^{\myx m/\mytau}$.
The density $n_m(r)$ is related to $f(\myx,m)$ as in  Eq.~(\ref{eq:spatial_density}), such that $n_m(r\ga r_h) \sim \myg (m)$.

\subsection{Negligible flow}
\label{sec:negligible_flow}
Consider the negligible flow limit $Q\to0$, which holds if the MF is not strongly dominated by light stars (AH09), as shown in \S\ref{sec:non_negligible_flow}.
Dividing Eq.~(\ref{eq:BW_masses}) by $m f(\myx,m)$ and
differentiating with respect to $m$ we find $\pr_m[(mf)^{-1}\pr_\myx
f]=0$, implying that $f$ must have the functional form
\begin{equation}
\label{eq:fm_form} f(\myx,m) = \myg (m)\myh(\myx)^m = \myg(m) e^{m \myk(\myx)} \coma
\end{equation}
with $\myk\equiv \ln \myh$.
Eq.~(\ref{eq:fm_form}) implies that the local power-law index of $f$, $\myp(\myx,m)\equiv d\ln f/d\ln \myx = m \myx \myk'(\myx)$, is linear in $m$ for fixed $\myx$. This result, valid for negligible $Q$, generalizes the BW77 result $\myp(\myx,m_1)/m_1=\myp(\myx,m_2)/m_2$ to a continuous MF.

Setting $Q=0$ in Eq.~(\ref{eq:BW_masses}), using Eq.~(\ref{eq:fm_form}),
and repeatedly differentiating with
respect to $\myx$ and isolating integral terms, yields an ODE which, for $q\neq0$, becomes
\begin{equation} \label{eq:ODE_masses}
B+(p/P)A = 0 \fin
\end{equation}
Here, $p$, $A$ and $B$ are functionals of $f(m,\myx)$ and $\myx$, defined as
in the equal mass case (\eg Eq.~(\ref{eq:ABC_defs})), and we defined
\begin{eqnarray}
\label{eq:average_P} P(\myx,m) & \equiv & m/4M_0 \coma \\
\label{eq:define_M0} M_L<M_0 & \equiv & \langle m^2 \rangle_f/\langle m \rangle_f = \langle m \rangle_{m f}<M_H \coma
\end{eqnarray}
with $\langle X \rangle_Y \equiv \int X(m) Y(m)\,dm/\int Y(m)\,dm$
averaging. The full distribution may be found by solving
Eq.~(\ref{eq:ODE_masses}) for $k(\myx)$ under the boundary conditions
$\myk(\myx\leq0)=\myx/\mytau$ and $\myk\to-\infty$ as $\myx\to\myx_{max}$.

Note that $P(\myx,M_L)\leq 1/4$ and $P(\myx,M_H)\geq 1/4$. If the mass range
is sufficiently narrow or $\myg $ is sufficiently dominated by high masses
such that $P(\myx,M_H)\simeq 1/4$, we recover the equal mass case
Eq.~(\ref{eq:ODEF0}) for the most massive stars. The full distribution then becomes
\begin{equation} \label{eq:fmassesp1_4}
f(\myx,m;\mytau) \simeq \myg (m) f_{eq}(x; \myT)^{m/M_H} \coma
\end{equation}
with $f_{eq}(x;\myT)$ the DF of equal-mass stars (Figure
\ref{fig:approx_single}).

\begin{figure}[h]
\centerline{\epsfxsize=7cm \epsfbox{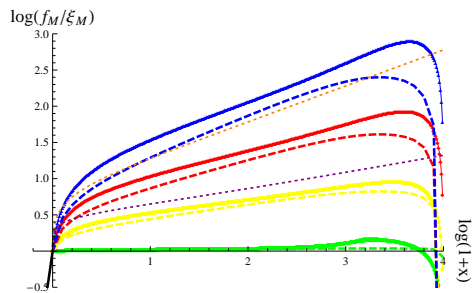}} \caption{
Unevolved Salpeter MF $\myg \propto M^{-2.35}$ with
$M_H/M_L=100$, $\mytau=(M_L M_H)^{1/2}$ and $\myx_{max}=10^4$.
DF shown (top to bottom) for $m/M_L=100, 67, 34$ and $1.5$, found by numerically solving Eq.~(\ref{eq:BW_masses}) (solid) and from Eq.~(\ref{eq:approx_masses}) (dashed). Also shown for reference are power law curves $6\myx^{1/2}$ and $2\myx^{1/4}$ (dotted).
}
\label{fig:approx_masses}
\end{figure}

\begin{figure}[h]
\centerline{\epsfxsize=10cm \epsfbox{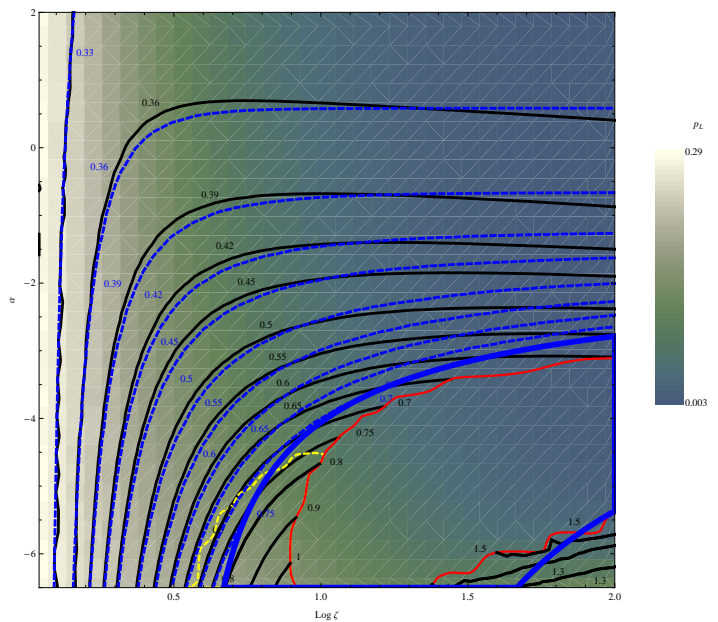}
} \caption{
Power-law MFs $\myg (M_L<m<M_H=\zeta M_L)\propto m^\alpha$ in the $\zeta-\alpha$ phase space for $\myx_{max}=10^4$ and $\mytau=(M_L M_H)^{1/2}$.
Shown are spectral indices $p_L,p_H$ of the lightest, heaviest (color scale, black contours) stars, found by numerically solving Eq.~(\ref{eq:BW_masses}). The power-law assumption $f(x,M_H)\propto \myx^{p_H}$ fails (p-value of $\chi^2$ fit larger than $1/2$, henceforth) in the region enclosed by red contours. The linear scaling $p\propto m$ breaks-down beneath the dashed yellow contour (wavy appearance of
these contours is a resolution effect).
Approximation Eq.~(\ref{eq:approx_pm2}) for $p_H$ is shown (dashed blue contours) outside the $\mydelta<0$ (see Eq.~(\ref{eq:determinant})) region, which is enclosed by thick blue contours.
}
\label{fig:p0}
\end{figure}

In the general case we may proceed as in \S\ref{sec:EqualMassStars}:
$f$ will be constant, vanish or diverge unless the two terms in
Eq.~(\ref{eq:ODE_masses}) are approximately balanced, \ie $p\simeq P$. If
$P$ depends only weakly on $\myx$, this implies a power-law DF, $f\propto \myx^{P}$. Thus, stars with mass equal to the weight-averaged mass $M_0$ tend to have a $p=1/4$
cusp as in the equal mass case. Taking some typical energy $\myx_{max}^{a}$ with $a\lesssim 1$ in Eq.~(\ref{eq:define_M0}), we may calibrate $a$  against numerical solutions of Eq.~(\ref{eq:BW_masses}). This yields $a\simeq1$ for a
wide range of MFs tested. Hence
\begin{equation} \label{eq:approx_pm2}
p(m) \simeq \myrho(\myx_{max})m/4M_0\coma \lrgspc M_0 = \langle m \rangle_{m \myg  \myx_{max}^{p(m)}} \coma
\end{equation}
and the full distribution is approximately
\begin{equation}
\label{eq:approx_masses} f (\myx,m; \mytau) \simeq \myg (m) f_{eq}\left( \myx;
\mytau \right)^{m/M_0} \fin
\end{equation}
Eqs.~(\ref{eq:approx_pm2}) and (\ref{eq:approx_masses}) agree (to better than a factor of $2$ in $f$) with numerical solutions of Eq.~(\ref{eq:BW_masses}) for various MFs, such as the discrete MFs tested by BW77 and AH09, the Salpeter MF (see Figure \ref{fig:approx_masses}) and a wide range of power-law MFs (see Figure \ref{fig:p0}).

Eq.~(\ref{eq:approx_pm2}) implies that for any MF, $p(m)$ is a decreasing function of $\myx_{max}$. In particular, the power-law index of the most massive
species, $p_H$, asymptotically approaches $1/4$ as $\myx_{max}\to\infty$, so $f$ approaches Eq.~(\ref{eq:fmassesp1_4}).
In cases where we may approximate $f\sim \myg $ in Eq.~(\ref{eq:define_M0}), for example
if $m\myg $ is strongly peaked in a mass range where $f/\myg =h^m$ varies little,
we have $M_0\simeq \langle m \rangle_{m\myg }$.

For concreteness, consider unbound stars with a
power-law MF $\myg (m)\propto m^\alpha$ in some mass range
$M_L<m<M_H=\myr M_L$. In this case we recover Eq.~(\ref{eq:ODE_masses}) with
\begin{equation} \label{eq:P_powerlaw_g}
P = -\frac{M\myk}{4} \,
\frac{\Gamma(2+\alpha,-M_H\myk,-M_L\myk)}
{\Gamma(3+\alpha,-M_H\myk,-M_L\myk)} \coma
\end{equation}
and $\Gamma(a,b,c)=\int_{b}^{c} t^{a-1}e^{-t}\,dt$ the incomplete
$\Gamma$-function. The power-law index $p(m)$ may be found from
Eq.~(\ref{eq:approx_pm2}),
\begin{equation} \label{eq:pH_power_law}
\frac
{\Gamma(3+\alpha,\frac{M_H}{m}\ln \myx_{max}^{-ap}, \frac{M_L}{m}\ln \myx_{max}^{-ap})}
{\Gamma(2+\alpha,\frac{M_H}{m}\ln \myx_{max}^{-ap}, \frac{M_L}{m}\ln \myx_{max}^{-ap})}
=\ln \myx_{max}^{-a/4}
\end{equation}
(for one mass, then using $p\propto m$).
Indices $p_L$, $p_H$ calculated from Eq.~(\ref{eq:pH_power_law}) are illustrated in
Figure \ref{fig:p0} in the $\myr-\alpha$ plane. For small (very negative) $\alpha$ we may estimate $p$ using the $k\to 0$ limit where $M_0\sim\langle m \rangle_{m\myg }$, whence
\footnote{Special cases: $\myp_H(\alpha=-2)\simeq \frac{\myr\ln \myr}{4(\myr-1)}$ and $\myp_H(\alpha=-3)\simeq \frac{\myr-1}{4\ln \myr}$.}
\begin{equation}
\myp_H \simeq
\frac{\myr(\alpha+3)(\myr^{2+\alpha}-1)}{4(\alpha+2)(\myr^{3+\alpha}-1)} \fin
\end{equation}

\subsection{Non-Negligible flow}
\label{sec:non_negligible_flow}
When $Q$ cannot be neglected, we may eliminate the $\myx'$ integral in Eq.~(\ref{eq:BW_masses}) by repeated operations of differentiation with respect to $\myx$ and isolation of integral terms. If we assume, in addition, a power-law energy dependence $f\simeq \myg (m)\myx^{p(m)}$,
we arrive at a generalization of Eq.~(\ref{eq:fixed_p0}),
\begin{eqnarray} \label{eq:p_vs_Q}
\myg (m) & \mathlarger{\int}_{M_L}^{M_H} &
\left[ m\left(p'+1\right)\left(p'-\frac{1}{2}\right)+m'p\left(\frac{3}{2}-p'\right) \right] \nonumber \\
& & \, \times \, m'\myg (m') \myx^{-\frac{3}{2}+p+p'} dm' \nonumber \\
& = & \frac{2}{3} p \left(p-1 \right) \left(p-\frac{5}{2}\right)Q(m) \coma
\end{eqnarray}
with abbrev. $p=p(m)$ and $p'=p(m')$. For $Q=0$ and $p\propto m$, this reproduces the results of \S \ref{sec:negligible_flow}, $p=m/4M_0$.

Eq.~(\ref{eq:p_vs_Q}) suggests that $p<3/2$ in all cases, otherwise the left hand side becomes large and strongly $\myx$-dependent.
Figure \ref{fig:p0} shows that the $p\simeq 3/2$ limit is indeed realized if the MF is sufficiently broad and light-dominated. It corresponds to mass segregation stronger even than the saturation value $p_H=5/4$ predicted by AH09 in the limit of a strongly bimodal, light-dominated MF where light stars are expected to assume $p_L=1/4$.

A variant of Eq.~(\ref{eq:p_vs_Q}) with $Q$ eliminated is
\begin{eqnarray} \label{eq:p_vs_Q_xi_indep}
& \mathlarger{\int}_{M_L}^{M_H} & dm' \myx^{p'}m'\myg (m') \left(\frac{3}{2}-p-p'\right) \\
& & \times \left[ m\left(p'+1\right)\left(p'-\frac{1}{2}\right)+m'p\left(\frac{3}{2}-p'\right) \right] = 0 \nonumber \fin
\end{eqnarray}
Adopting a typical specific energy, $\myx_{max}^b$ with $b\lesssim 1$,
one can solve this equation for $p(m)$ with arbitrary $\myg (m)$, an approximate procedure far simpler than solving Eq.~(\ref{eq:BW_masses}). However, Eq.~(\ref{eq:p_vs_Q_xi_indep}) typically becomes unstable when the underlying power-law assumption
$f\sim \myx^p$ fails.

The linear scaling $p(m)\propto m$ remains approximately valid even when the flow is only marginally negligible. Using this Ansatz in Eq.~(\ref{eq:p_vs_Q_xi_indep}), $p(m)$ becomes a root of the quadratic equation
\begin{equation} \label{eq:Qapprox}
4p^2 \langle m^2 \rangle - p(7-4p) m \langle m \rangle + \left(3/2-p\right)m^2=0 \coma
\end{equation}
with averages weighted by $m \myg  \myx_{max}^{bp}$.
For light-dominated MFs and massive stars with $m\gg\{\langle m \rangle, \langle m^2 \rangle^{1/2} \}$, the last term dominates and $p(m)$
peaks at $3/2$. For a strongly peaked MF where $\langle m^2 \rangle \simeq \langle m \rangle^2$, Eq.~(\ref{eq:Qapprox}) yields two solutions: the $Q=0$ limit $p=m/4\langle m \rangle$, and a steep solution $p=(3/2)(1+\langle m \rangle/m)^{-1}$.

Eq.~(\ref{eq:Qapprox}) suggests the following, numerically supported picture. For MFs peaked at relatively massive stars, the DF is approximately given by the $Q=0$ result. For more light-dominated MFs, $M_0$ is shifted towards
lower masses so the DF becomes gradually steeper at any given mass. If the MF is sufficiently extended (large $\myr$) and light-dominated (\eg small $\alpha$), each stellar species of mass $m\gg M_L$ eventually achieves maximal steepness $p(m)\simeq 3/2$. The transition between the $Q=0$ and the saturation regimes involves in general an intermediate range of parameters in which the DF of this species is no longer a power law in $\myx$. We may roughly identify this region with the absence of real solutions to Eq.~(\ref{eq:Qapprox}),
\begin{equation} \label{eq:determinant}
\mydelta(m)=(m/\langle m \rangle-5)^2-24(\langle m^2 \rangle/\langle m \rangle^2-1) <0 \coma
\end{equation}
illustrated as areas delimited by solid curves in Figure \ref{fig:p0}. After reaching maximal steepness $f(m)$ continues to evolve, but its maximal value remains $\sim f(x=1)x_{max}^{3/2}$.

\section{Conclusions}

We have analyzed the distribution of an arbitrary, continuous MF $\myg $ of stars around a MBH. For equal mass stars, we derive a simple approximation for the DF $f$, see Eq.~(\ref{eq:approx_single}) and Figure \ref{fig:approx_single}. For MFs not strongly dominated by light stars, where the mass flow can be neglected, we generalize the BW77 linear scaling $p\propto m$ to continuous MFs. We then derive approximate solutions for $p$ (including its normalization, Eq.~(\ref{eq:approx_pm2})) and $f$ (Eq.~(\ref{eq:approx_masses})) and confirm them numerically; see Figures \ref{fig:approx_masses} and \ref{fig:p0}.

Our results provide a simple yet accurate alternative to solving the full integro-differential equation (\ref{eq:BW_masses}).
Eq.~(\ref{eq:approx_pm2}) reproduces and generalizes previous Fokker-Planck calculations such as BW77, which were limited to a narrow range of parameters. As an illustration, consider SgA* with $x_{max}=10^4$ and a model MF with main sequence stars, white dwarfs, neutron stars and black holes, of masses $M/M_\odot=1,0.6,1.4,10$, and relative abundances $1:10^{-1}:10^{-2}:10^{-3}$. Eq.~(\ref{eq:approx_pm2}) then gives $M_0=4.8M_\odot$ so $p_H=0.52$ for black holes and $p<0.08$ for the other species, in agreement with the numerical results of \citet{HopmanAlexander06} and AH09.

The DF becomes gradually steeper with decreasing $M_0$, as long as $\mydelta$ in Eq.~(\ref{eq:determinant}) remains positive. For very light-dominated MFs, the power law assumption $f\sim \myg \myx^p$ eventually fails at high masses, roughly where $\mydelta(m)<0$ (contour-enclosed regions in Figure \ref{fig:p0}). As the MF peak shifts to even lower masses, $\mydelta(m)>0$ and the power-law behavior are eventually restored at the high mass end and the DF remains very steep, peaking at $p=3/2$. This behavior is illustrated in Figure \ref{fig:p0}.

Power law DFs $f\propto \myx^{p}$ correspond to $n(r)\propto r^{-3/2-p}$. As long as the MF is not very light-dominated, stars of mass $M_0$ have $p_0=1/4$ (up to a $\myrho$ correction) and an
equal-mass like cusp $n_0\propto r^{-7/4}$, while other stars have
$n_m/n_0 \propto r^{-(m-M_0)/4M_0}$. If the MF is broad and heavy-dominated, light stars tend towards an energy independent distribution $n\propto r^{-3/2}$ while heavy stars approach $n\propto r^{-7/4}$. In the opposite, maximally steep limit, heavy stars develop a cusp as steep as $n\propto r^{-3}$, harder than any reported previously, and the number of stars in the cusp depends (logarithmically) on the disruption cutoff.

\acknowledgements We thank G. Van de Ven, D. Keres and Y. Birnboim for helpful
discussions. The research of U.K. was supported in part by NSF grant PHY-0503584, as a Friends of the Institute for Advanced Study member. The research of
C.H. is supported by a Veni fellowship from the Netherlands
Organization for Scientific Research (NWO). C.H. thanks the
hospitality of the Institute for Advanced Study, where part of this
research took place.
T.A. is supported by ISF grant
928/06 and ERC grant 202996.

\bibliographystyle{apsrev}

\end{document}